\documentclass[12pt]{article}
\usepackage{amsmath}
\usepackage{amssymb}
\begin{document}
\setcounter{page}{1}
\def\theequation{\arabic{section}.\arabic{equation}}
\def\theequation{\thesection.\arabic{equation}}
\setcounter{section}{0}

\title{Singularities of ultra--relativistic expanding quark--gluon
plasma within Quark--gluon transport theory}

\author{A. Ya. Berdnikov , Ya. A. Berdnikov~\thanks{E--mail:
berdnikov@twonet.stu.neva.ru} ,\\ A. N. Ivanov , V. A. Ivanova ,
V. F. Kosmach ,\\ V. M. Samsonov~\thanks{E--mail:
samsonov@hep486.pnpi.spb.ru, St. Petersburg Institute for Nuclear
Research, Gatchina, Russian Federation} , N. I. Troitskaya}

\date{\today}

\maketitle

\begin{center}
{\it Department of Nuclear Physics, State Technical University of
St. Petersburg,\\ 195251 St. Petersburg, Russian Federation}
\end{center}

\begin{center}
\begin{abstract}
We follow Quark--gluon transport theory and analyse singularities of
the ultra--relativistic and spherical expanding quark--gluon plasma.
Within the {\it linearized} QCD oscillations and instabilities of the
ultra--relativistic and spherical symmetric expanding quark--gluon
plasma near global thermodynamical equilibrium are investigated in
dependence on a chemical potential $\mu(T)$ of non--strange light
quarks and antiquarks, a strange quark mass $m_s$, a temperature $T$
and a hydrodynamical velocity $u$. We calculate the chromoelectric
permeability tensor for the quark--gluon plasma at rest and
ultra--relativistic moving. We show that the contribution of chemical
potential of light quarks and antiquarks can be neglected to the
chromoelectric permeability. We show that the account for the
non--zero mass of strange quarks and antiquarks diminishes the value
of the plasma frequency. We show that the plasma frequency of the
ultra--relativistic and spherical symmetric expanding quark--gluon
plasma is enhanced by a Lorentz factor compared with the plasma
frequency of the quark--gluon plasma at rest. We find that the
ultra--relativistic and spherical symmetric expanding quark--gluon
plasma behaves like a collisionless thermalized plasma.\\
\vspace{0.2in}
\noindent PACS numbers: 25.75.--q, 12.38.Mh, 24.85.+p
\end{abstract}
\end{center}

\newpage

\section{Introduction}
\setcounter{equation}{0}

\hspace{0.2in} One of the most meaningful and challenging problems of
the modern high--energy physics is theoretical and experimental
investigation of the quark--gluon plasma (QGP) [1]. There is a believe
that the QGP phase of the quark--gluon system can be realized in
ultra--relativistic heavy--ion collisions [1]. In this paper we study
the problem of theoretical description of singularities of the QGP,
such as oscillations and instabilities, produced in the intermediate
state of ultra--relativistic heavy--ion collisions in the center of
mass frame of colliding ions. In our approach to the description of
the QGP we follow the quark--gluon transport (QGT) theory outlined in
[2--4] and based on the relativistic kinetic theory nicely expounded
in the book [5].

The main objects of the QGT approach are one--particle distribution
functions $f_q(x,p)$, $\bar{f}_{\bar{q}}(x,p)$ and $f_g(x,p)$ defining
probabilities to find a quark $q = u, d$ or $s$, an antiquark $\bar{q} =\bar{u}, \bar{d}$ or $\bar{s}$ and a gluon
$g$ at a space--time point $x = x^{\mu} = (t,\vec{x}\,)$ with a
4--momentum $p = p^{\mu} = (E, \vec{p}\,)$, respectively [2--4].  The
distribution functions of quarks, $f_q(x,p)$, and antiquarks,
$\bar{f}_{\bar{q}}(x,p)$, are hermitian $N_C\times N_C$ matrices,
whereas the gluon distribution function $f_g(x,p)$ is a hermitian
$(N^2_C-1)\times (N^2_C-1)$ matrix in colour space of a $SU(N_C)$
colour gauge group. These distribution functions obey the transport
equations [2,3,6]
\begin{eqnarray}\label{label1.1}
p^{\mu}D_{\mu}f_u(x,p) +
\frac{1}{2}\,g_s\,p^{\mu}\frac{\partial}{\partial
p_{\nu}}\{G_{\mu\nu}(x),f_u(x,p)\} &=&
C_u[f_u,f_d,f_s,\bar{f}_{\bar{u}},\bar{f}_{\bar{d}},
\bar{f}_{\bar{s}},f_g],\ldots ,\nonumber\\
p^{\mu}D_{\mu}\bar{f}_{\bar{u}}(x,p) -
\frac{1}{2}\,g_s\,p^{\mu}\frac{\partial}{\partial
p_{\nu}}\{G_{\mu\nu}(x),\bar{f}_{\bar{u}}(x,p)\}
&=&\bar{C}_{\bar{u}}[f_u,f_d,f_s,\bar{f}_{\bar{u}},\bar{f}_{\bar{d}},
\bar{f}_{\bar{s}},f_g],\ldots,\nonumber\\ p^{\mu}{\cal
D}_{\mu}f_g(x,p) + \frac{1}{2}\,g_s\,p^{\mu}\frac{\partial}{\partial
p_{\nu}}\{{\cal G}_{\mu\nu}(x),f_g(x,p)\} &=&
C_g[f_u,f_d,f_s,\bar{f}_{\bar{u}},\bar{f}_{\bar{d}},
\bar{f}_{\bar{s}},f_g],
\end{eqnarray}
where $g_s$ is a quark--gluon coupling constant, the symbol
$\{\ldots,\ldots\}$ denotes anticommutators, $D_{\mu}$ and ${\cal
D}_{\mu}$ are the covariant derivatives in the fundamental and adjoint
representations,respectively, which act as
\begin{eqnarray}\label{label1.2}
D_{\mu} &=& \partial_{\mu} - i\,g_s\,[A_{\mu}(x),\ldots],\nonumber\\
{\cal D}_{\mu} &=& \partial_{\mu} - i\,g_s\,[{\cal A}_{\mu}(x),\ldots].
\end{eqnarray}
The gluon field potentials $A_{\mu}(x)$ and ${\cal A}_{\mu}(x)$ are
determined by
\begin{eqnarray}\label{label1.3}
A_{\mu}(x) = A^a_{\mu}t^a_C\quad,\quad {\cal A}_{\mu}(x) = {\cal
A}^a_{\mu}(x)T^a_C.
\end{eqnarray}
The matrices $t^a_C$ and $T^a_C\,(a=1,2,\ldots,N^2_C-1)$ are the
generators of the $SU(N_C)$ gauge group in the fundamental,
$(t^a_C)_{ij} = (\lambda^a_C)_{ij}/2$\,\footnote{Here $\lambda^a_C$
are Gell--Mann's matrices of the $SU(N_C)$ group.} with indices
$i\,(j)$ running over $1,2,\ldots,N_C$, and adjoint, $(T^a_C)^{bc} =
-\,i\,f^{abc}$ with indices $a\,(b,c)$ running over
$1,2,\ldots,N^2_C-1$, representations, where $f^{abc}$ are the
structure constants of the $SU(N_C)$ gauge group defined by
\begin{eqnarray}\label{label1.4}
[X^a_C,X^b_C] = i\,f^{abc}\,X^c_C,
\end{eqnarray}
where $X^a_C$ is either $t^a_C$ or $T^a_C$.  The matrices $t^a_C$ and
 $T^a_C$ are normalized by the conditions ${\rm tr}_C\{t^at^b\} =
 \delta^{ab}/2$ and ${\rm tr}_C\{T^aT^b\} = N_C\,\delta^{ab}$. Then,
 $G_{\mu\nu}(x)$ and ${\cal G}_{\mu\nu}(x)$ are the gluon mean--field
 strength tensors in the fundamental and adjoint representations
\begin{eqnarray}\label{label1.5}
G_{\mu\nu}(x) &=&\partial_{\mu}A_{\mu}(x) - \partial_{\nu}A_{\mu}(x)
-\,i\,g\,[A_{\mu}(x), A_{\nu}(x)],\nonumber\\ {\cal G}_{\mu\nu}(x)
&=&\partial_{\mu}{\cal A}_{\mu}(x) - \partial_{\nu}{\cal A}_{\mu}(x)
-\,i\,g_s\,[{\cal A}_{\mu}(x), {\cal A}_{\nu}(x)].
\end{eqnarray}
The gluon mean--field is
induced by the current of quarks, antiquarks and gluons [3]
\begin{eqnarray}\label{label1.6}
D_{\mu}G^{\mu\nu}(x) = \partial_{\mu}G^{\mu\nu}(x)
-i\,g_s\,[A_{\mu}(x),G^{\mu\nu}(x)] = j^{\nu}(x),
\end{eqnarray}
where the current $j^{\nu}(x)$ is determined by quark, antiquark
and gluon distribution functions
\begin{eqnarray}\label{label1.7}
&&j^{\mu}(x) = -\,\frac{1}{2}\,g_s\int\frac{d^3p}{(2\pi)^3
E_{\vec{p}}}\,p^{\mu}\Big\{\sum_{q =u,d,s}g_q\,f_q(x,p) -
\sum_{\bar{q} =
\bar{u},\bar{d},\bar{s}}g_{\bar{q}}\,\bar{f}_{\bar{q}}(x,p)\nonumber\\
&& - \frac{1}{N_C}\,{\rm tr}_C[\sum_{q =u,d,s}g_q\,f_q(x,p) -
\sum_{\bar{q} =
\bar{u},\bar{d},\bar{s}}g_{\bar{q}}\,\bar{f}_{\bar{q}}(x,p)] +
2\,g_g\,i\,t^a_C\,f^{abc}\,f_g(x,p)^{bc}\Big\},
\end{eqnarray}
where $g_q$, $g_{\bar{q}}$ and $g_g$ are spin degeneracies of quarks,
antiquarks and gluons. The transport equations (\ref{label1.1})
supplemented by equations (\ref{label1.6}) and (\ref{label1.7})
represent a complete set of relativistic kinetic equations for a
non--equilibrium quark--gluon system in Vlasov's approach [5].

Equations (\ref{label1.1}), (\ref{label1.6}) and (\ref{label1.7})
should be considered together with colour independent quantities [3]
which are (i) the baryon number current $b^{\mu}(x)$
\begin{eqnarray}\label{label1.8}
b^{\nu}(x) = \int\frac{d^3p}{(2\pi)^3 E_{\vec{p}}}\,p^{\mu}\,{\rm
tr}_C[\sum_{q =u,d,s}g_q\,f_q(x,p) - \sum_{\bar{q} =
\bar{u},\bar{d},\bar{s}}g_{\bar{q}}\,\bar{f}_{\bar{q}}(x,p)]
\end{eqnarray}
and (ii) the energy--momentum tensor $t^{\mu\nu}(x)$
\begin{eqnarray}\label{label1.9}
t^{\mu\nu}(x) = \int\frac{d^3p}{(2\pi)^3
E_{\vec{p}}}\,p^{\mu}p^{\nu}\,{\rm tr}_C[\sum_{q =u,d,s}g_q\,f_q(x,p)
+ \sum_{\bar{q} =
\bar{u},\bar{d},\bar{s}}g_{\bar{q}}\,\bar{f}_{\bar{q}}(x,p) +
g_g\,f_g(x,p)].
\end{eqnarray}
The r.h.s. of (\ref{label1.1}) is defined by the collision terms $C_u,
C_d, C_s$, $\bar{C}_{\bar{u}},\bar{C}_{\bar{d}},\bar{C}_{\bar{s}}$ and
$C_g$. These collision terms vanish in the collisionless limit, when
the gluon mean--field effects dominate for the evolution of the
QGP. Such a dominance appears when the characteristic frequency of
variations of a gluon mean--field becomes much greater than a frequency
of parton collisions [3]. 

As has been emphasized by Mr${\acute{\rm o}}$wczy${\acute{\rm n}}$ski
in Ref.[3] the collision terms of the QGP kinetic equations
(\ref{label1.1}) have not been jet derived, but they can be taken in
the form of the collision terms of the Waldmann--Snider kinetic
equations [3,7] describing a non--equilibrium system of spinning
particles [5]. Recall, that the Waldmann--Snider kinetic equations
play an important role in the kinetic theory of multicomponent atomic
gases [5,8].

The collision terms in the relaxation time approximation can be
approximated by [3]
\begin{eqnarray*}
C_u[f_u,f_d,f_s,\bar{f}_{\bar{u}},\bar{f}_{\bar{d}},
\bar{f}_{\bar{s}},f_g]&=&\nu_u\,p_{\mu}u^{\mu}(x)\,[f^{\rm eq}_u(x,p)
- f_u(x,p)],\nonumber\\
C_d[f_u,f_d,f_s,\bar{f}_{\bar{u}},\bar{f}_{\bar{d}},
\bar{f}_{\bar{s}},f_g]&=&\nu_d\,p_{\mu}u^{\mu}(x)\,[f^{\rm eq}_d(x,p)
- f_d(x,p)],\nonumber\\
C_s[f_u,f_d,f_s,\bar{f}_{\bar{u}},\bar{f}_{\bar{d}},
\bar{f}_{\bar{s}},f_g]&=&\nu_s\,p_{\mu}u^{\mu}(x)\,[f^{\rm eq}_s(x,p)
- f_s(x,p)],\nonumber\\
\bar{C_{\bar{u}}}[f_u,f_d,f_s,\bar{f}_{\bar{u}},\bar{f}_{\bar{d}},
\bar{f}_{\bar{s}},f_g] &=&
\nu_{\bar{u}}\,p_{\mu}u^{\mu}(x)\,[\bar{f}^{\rm eq}_{\bar{u}}(x,p) -
\bar{f}_{\bar{u}}(x,p)],\nonumber\\
\bar{C_{\bar{d}}}[f_u,f_d,f_s,\bar{f}_{\bar{u}},\bar{f}_{\bar{d}},
\bar{f}_{\bar{s}},f_g] &=&
\nu_{\bar{d}}\,p_{\mu}u^{\mu}(x)\,[\bar{f}^{\rm eq}_{\bar{d}}(x,p) -
\bar{f}_{\bar{d}}(x,p)],
\end{eqnarray*}
\begin{eqnarray}\label{label1.10}
\bar{C_{\bar{s}}}[f_u,f_d,f_s,\bar{f}_{\bar{u}},\bar{f}_{\bar{d}},
\bar{f}_{\bar{s}},f_g] &=&
\nu_{\bar{s}}\,p_{\mu}u^{\mu}(x)\,[\bar{f}^{\rm eq}_{\bar{s}}(x,p) -
\bar{f}_{\bar{s}}(x,p)],\nonumber\\
C_g[f_u,f_d,f_s,\bar{f}_{\bar{u}},\bar{f}_{\bar{d}},
\bar{f}_{\bar{s}},f_g] &=& \nu_g\,p_{\mu}u^{\mu}(x)[f^{\rm eq}_g(x,p)
- f_g(x,p)],
\end{eqnarray}
where $\nu_q\,(q = u,d,s)$,
$\nu_{\bar{q}}\,(\bar{q}=\bar{u},\bar{d},\bar{s})$ and $\nu_g$ are the
equilibration rate parameters, the inverse relaxation times, which can
be identified with inverse free flight times for quarks, antiquarks
and gluons [3]; $u^{\mu}(x)$ is a hydrodynamical 4--velocity. Then,
$f^{\rm eq}_q(x,p)$, $\bar{f}^{\rm eq}_{\bar{q}}(x,p)$ and $f^{\rm
eq}_g(x,p)$ are the equilibrium distribution functions.  These are
Fermi--Dirac distribution functions for quarks and antiquarks
\begin{eqnarray}\label{label1.11}
f^{\rm eq}_q(x,p)_{ij} &=& \delta_{ij}\,n_q(p,T) =\delta_{ij}\,
\frac{1}{\displaystyle e^{\textstyle (u^{\mu}p_{\mu} - \mu(T))/T}
+1},\nonumber\\ \bar{f}^{\rm eq}_{\bar{q}}(x,p)_{ij} &=&\delta_{ij}\,
n_{\bar{q}}(p,T) = \delta_{ij}\,\frac{1}{\displaystyle e^{\textstyle
(u^{\mu}p_{\mu} + \mu(T))/T} +1},
\end{eqnarray}
where $T$ is a temperature measured in MeV and $\mu(T)$ is a chemical
potential of massless quarks, and the Bose--Einstein distribution
function for gluons 
\begin{eqnarray}\label{label1.12}
f^{\rm eq}_g(x,p)_{ab} = \delta_{ab}\,n_g(p,T) =
\delta_{ab}\,\frac{1}{\displaystyle e^{\textstyle u^{\mu}p_{\mu}/T} -
1}.
\end{eqnarray}
Neglecting the contribution of statistics, i.e. unities with respect
to exponentials in Eqs.(\ref{label1.11}) and (\ref{label1.12}),
distribution functions (\ref{label1.11}) and (\ref{label1.12}) reduce
themselves to the form introduced by J\"uttner [9].

As has been discussed in details in Refs.[2,3] such a transport
approach to the QGP allows to analyse singular properties of the QGP
such as oscillations and instabilities. The main aim of this paper is
to analyse oscillations and instabilities of the QGP in dependence on
a chemical potential of light $u$ and $d$ quarks $\mu(T)$ as a
function of a temperature $T$, a mass $m_s$ of strange quark quarks
and antiquarks and a hydrodynamical velocity $u^{\mu}$. 

In the electromagnetic plasma such phenomena as oscillations and
instabilities can be described in terms of singularities of the
electric permeability tensor $\varepsilon^{ij}(k)$, where Latin
indices run over $i = 1,2,3$ and $k^{\mu} = (\omega,\vec{k}\,)$ is a
4--momentum [10]. As usual the electric permeability tensor is
expanded into transversal and longitudinal parts with respect to a
3--momentum $\vec{k}$ [10]
\begin{eqnarray}\label{label1.13}
\varepsilon^{ij}(k) =
\varepsilon_T(\omega,\vec{k}\,)\,\Bigg(\delta^{ij} -
\frac{k^ik^j}{\vec{k}^{\,2}}\Bigg) +
\varepsilon_L(\omega,\vec{k}\,)\,\frac{k^ik^j}{\vec{k}^{\,2}}.
\end{eqnarray}
The calculation of $\varepsilon_T(\omega,\vec{k}\,)$ and
$\varepsilon_L(\omega,\vec{k}\,)$ depends on the model describing a
dynamics of plasma constituents [10].

As has been pointed out in Refs.[2,3] the calculation of the
chromoelectric permeability tensor within QCD is achievable only for
the {\it linearized} QCD. In the {\it linearized} QCD and for the QGP
near global thermodynamical equilibrium oscillations and instabilities
of the QGP can be described in terms of the chromoelectric
permeability tensor $\varepsilon^{ij}(k)$ determined by [3]
\begin{eqnarray}\label{label1.14}
\varepsilon^{ij}(k) = \delta^{ij} - \frac{i}{\omega}\,\sigma^{i 0
j}(k) - \frac{i}{\omega^2}\,k^{\ell}\,[\sigma^{i j \ell}(k) -
\sigma^{i \ell j}(k)],
\end{eqnarray}
where $\sigma^{\alpha\beta\lambda}(k)$ is the colour conductivity
tensor [3].

The paper is organized as follows. In section 2 we discuss the
calculation of the colour conductivity tensor
$\sigma^{\alpha\beta\lambda}(k)$. We write down the components of the
colour conductivity tensor in dependence on a chemical potential of
light non--strange quarks and antiquarks, a mass of strange quarks and
a hydrodynamical velocity of the QGP. In section 3 we calculate the
chromoelectric permeability tensor for the QGP at rest.  We show that
the contribution of a chemical potential of light quarks and
antiquarks can be neglected, but the contribution of the non--zero
mass of strange quarks and antiquarks leads to screening of strange
quarks and antiquarkes for the formation of the chromoelectric
permeability. This entails the decrease of the plasma frequency
calculated in Refs.[2,3]. The numerical estimate is carried out for
$m_s = 135\,{\rm MeV}$ and $T = 175\,{\rm MeV}$. In section 4 we
calculate the chromoelectric permeability tensor for the
ultra--relativistic and spherical symmetric expanding QGP. We show
that the chromoelectric permeability tensor retains its form
calculated for the QGP at rest save the value of the plasma frequency
which becomes enhanced by a twice Lorentz factor of the
ultra--relativistic motion of the QGP. In the Conclusion we discuss
the obtained results. We argue that our results are in agreement with
those obtained by Elze and Heinz [2] and Mr${\acute{\rm
o}}$wczy${\acute{\rm n}}$ski [3]. Analytical expressions for the
solutions of dispersion relations derived in Refs.[2,3] for
singularities of the QGP defining oscillations and instabilities are
fully retained. The distinctions of our results from those of
Refs.[2,3] concern the magnitudes of the plasma frequency which is
slightly diminished for the QGP at rest due to screening of strange
quarks and antiquarks and substantially enhanced by a twice Lorentz
factor for the ultra--relativistic and spherical symmetric expanding
QGP.

\section{Colour conductivity tensor of quark--gluon plasma}
\setcounter{equation}{0}

\hspace{0.2in} The colour conductivity tensor
$\sigma^{\alpha\beta\lambda}(k)$, the components of which define the
chromoelectric permeability Eq.(\ref{label1.14}) and derived in
Ref.[3], can be transcribed as follows
\begin{eqnarray}\label{label2.1}
\hspace{-0.5in}&&\sigma^{\alpha\beta\lambda}(k) =
i\,g^2_s\int\frac{d^3p}{(2\pi)^3}\,
\frac{p^{\alpha}p^{\beta}}{E_{\vec{p}}}\,
\Bigg[\sum_{q=u,d}\frac{1}{p\cdot (k + i\nu_{q}u)}\,\frac{\partial
n_q(p,T)}{\partial p_{\lambda}} + \frac{1}{p\cdot (k +
i\nu_{s}u)}\frac{\partial n_s(p,T)}{\partial p_{\lambda}}\nonumber\\
\hspace{-0.5in}&& + \sum_{\bar{q}=\bar{u},\bar{d}}\frac{1}{p\cdot (k +
i\nu_{\bar{q}} u)}\frac{\partial n_{\bar{q}}(p,T)}{\partial
p_{\lambda}} + \frac{1}{p\cdot (k + i\nu_{\bar{s}}u)}\frac{\partial
n_{\bar{s}}(p,T)}{\partial p_{\lambda}} + \frac{2\,N_C\,}{p\cdot (k +
i\nu_{g}u)}\frac{\partial n_g(p,T)}{\partial p_{\lambda}}\Bigg],
\end{eqnarray}
where we have set $g_q = g_{\bar{q}} = g_g = 2$.  For the calculation
of the colour conductivity tensor Eq.(\ref{label2.1}) we would use the
probabilities of the light quarks and antiquarks defined by
Eq.(\ref{label1.11}) and the probabilities of strange quarks and
antiquarks given by
\begin{eqnarray}\label{label2.2}
n_s(p,T) = n_{\bar{s}}(p,T) = \frac{1}{\displaystyle e^{\textstyle
u^{\mu}p_{\mu}/T} + 1}.
\end{eqnarray}
At the rest frame when $u^{\mu} = (1,\vec{0}\,)$ the probabilities of
strange quarks and antiquarks read
\begin{eqnarray}\label{label2.3}
n_s(\vec{p}, T) = n_{\bar{s}}(\vec{p}, T) = \frac{1}{\textstyle
e^{\textstyle \sqrt{\vec{p}^{\,\,2} + m^2_s}/T} + 1},
\end{eqnarray}
where $m_s = 135\,{\rm MeV}$ [11] is the mass of the strange quark and
antiquark. The value of the current $s$--quark mass $m_s =135\,{\rm
MeV}$ has been successfully applied to the calculation of chiral
corrections to the amplitudes of low--energy interactions, form
factors and mass spectra of low--lying hadrons [12] and charmed
heavy--light mesons [13]. Unlike massless antiquarks $\bar{u}$ and
$\bar{d}$ for which a low--temperature suppression, $T \to 0$, is
caused by a chemical potential $\mu(T)$, strange quarks and antiquarks
are suppressed at $T \to 0$ by virtue of a non--zero mass $m_s$.

For the QGP, realized for ultra--relativistic heavy--ion collisions, a
hydrodynamical velocity $u$ we define as
\begin{eqnarray}\label{label2.4}
u^{\mu} = (\gamma,\gamma\,\vec{v}\,) = \Bigg(\frac{1}{\sqrt{1
- v^2}}, \frac{\vec{v}}{\sqrt{1 - v^2}}\Bigg),
\end{eqnarray}
where $\vec{v}$ is a 3--dimensional hydrodynamical velocity of the QGP
such as $v \sim 1$. This gives the constraint $\gamma \gg 1$.

In order to derive the colour conductivity tensor given by
Eq.(\ref{label2.1}) one has to calculate the deviations of the
distribution functions of the QGP constituents from their equilibrium
distribution functions [3]. For the QGP the {\it coloured} state of
which is close to the quasi--stable {\it colourless} homogeneous state
the distribution functions for quarks, antiquarks and gluons can be
defined by [3]
\begin{eqnarray}\label{label2.5}
f_q(x,p)_{ij} &=& n_q(p)\,\delta_{ij} + \delta
f_q(x,p)_{ij},\nonumber\\ \bar{f}_{\bar{q}}(x,p)_{ij} &=&
\bar{n}_{\bar{q}}(p)\,\delta_{ij} + \delta
\bar{f}_{\bar{q}}(x,p)_{ij},\nonumber\\ f_g(x,p)_{ab} &=&
n_g(p)\,\delta_{ab} + \delta f_g(x,p)_{ab}.
\end{eqnarray}
The functions $\delta f_q(x,p)_{ij}$, $\delta \bar{f}_{\bar{q}}(x,p)$
and $\delta f_g(x,p)_{ab}$ describing deviations from the {\it
colourless} equilibrium state as well as gradients of these functions
are smaller compared with the equilibrium distribution functions
[3]. Then, traces of these functions vanish, i.e. ${\rm
tr}_C\{f_q(x,p)\} = {\rm tr}_C\{\bar{f}_{\bar{q}}(x,p)\} = {\rm
tr}_C\{f_g(x,p)\} = 0$.

Substituting Eq.(\ref{label2.5}) into Eq.(\ref{label1.7}) we obtain
\begin{eqnarray}\label{label2.6}
\hspace{-0.1in}&&j^{\mu}(x) = -\frac{g_s}{2}\int\frac{d^3p}{(2\pi)^3
}\frac{p^{\mu}}{E_{\vec{p}}}\nonumber\\
\hspace{-0.1in}&&\times\,\Big\{\sum_{q =u,d,s}g_q\,\delta f_q(x,p) -
\sum_{\bar{q} = \bar{u},\bar{d},\bar{s}}g_{\bar{q}}\,\delta
\bar{f}_{\bar{q}}(x,p) + 2\,g_g\,i\,t^a_C\,f^{abc}\,\delta
f_g(x,p)^{bc}\Big\}.
\end{eqnarray}
According to [3] the distribution functions $\delta f_q(x,p)$, $\delta
\bar{f}_{\bar{q}}(x,p)$ and $\delta f_g(x,p)$ can be obtained as
solutions of the linearized Eq.(\ref{label1.1}) with collision terms
given by Eq.(\ref{label1.10}). The corresponding solutions read [3]
\begin{eqnarray}\label{label2.7}
\delta f_q(x,p) &=& -ig_s\int
\frac{d^4k}{(2\pi)^4}\,\frac{\displaystyle e^{\textstyle -ik\cdot
x}}{p\cdot(k + i\nu_q u)}\,p^{\beta}G_{\beta\nu}(k)\,\frac{\partial
n_q(p,T)}{\partial p_{\nu}},\nonumber\\ \delta \bar{f}_{\bar{q}}(x,p)
&=& +ig_s\int \frac{d^4k}{(2\pi)^4}\,\frac{\displaystyle e^{\textstyle
-ik\cdot x}}{p\cdot(k +
i\nu_{\bar{q}}u)}\,p^{\beta}G_{\beta\nu}(k)\,\frac{\partial
n_{\bar{q}}(p,T)}{\partial p_{\nu}},\nonumber\\ \delta f_g(x,p) &=&
-ig_s\int \frac{d^4k}{(2\pi)^4}\,\frac{\displaystyle e^{\textstyle
-ik\cdot x}}{p\cdot(k + i\nu_g u)}\,p^{\beta}{\cal
G}_{\beta\nu}(k)\,\frac{\partial n_g(p,T)}{\partial p_{\nu}},
\end{eqnarray}
where $n_q(p,T)$, $n_{\bar{q}}(p,T)$ and $n_g(p,T)$ are equilibrium
distribution functions of quarks, antiquarks and gluons, respectively.
Then, $G_{\beta\nu}(k)$ and ${\cal G}_{\beta\nu}(k)$ are Fourier
transforms of the gluon mean--field strength tensors in the fundamental
and adjoint representations, respectively, such as
\begin{eqnarray}\label{label2.8}
{\cal G}^{ab}_{\beta\nu}(k) = -if^{abc}G^c_{\beta\nu}(k).
\end{eqnarray}
Substituting Eq.(\ref{label2.7}) in Eq.(\ref{label2.6}) we get
\begin{eqnarray}\label{label2.9}
\hspace{-0.3in}&&j^{\alpha}(x) = i\,\frac{g^2_s}{2}\int
\frac{d^4k}{(2\pi)^4}\,e^{\textstyle -ik\cdot
x}\int\frac{d^3p}{(2\pi)^3
}\frac{p^{\alpha}p^{\beta}\,}{E_{\vec{p}}}\Bigg[\sum_{q
=u,d,s}\frac{g_q}{p\cdot(k + i\nu_qu)}\,\frac{\partial
n_q(p,T)}{\partial p_{\lambda}}\nonumber\\ 
\hspace{-0.3in}&&+ \sum_{\bar{q} =
\bar{u},\bar{d},\bar{s}}\frac{g_{\bar{q}}}{p\cdot(k +
i\nu_{\bar{q}}u)}\,\frac{\partial n_{\bar{q}}(p,T)}{\partial
p_{\lambda}}\delta \bar{f}_{\bar{q}}(x,p) + \frac{2N_Cg_g}{p\cdot(k +
i\nu_gu)}\,\frac{\partial n_g(p,T)}{\partial
p_{\lambda}}\Bigg]\,G_{\beta\nu}(k) = \nonumber\\ \hspace{-0.3in}&& =
\int \frac{d^4k}{(2\pi)^4}\,j^{\alpha}(k)\,e^{\textstyle -ik\cdot x} =
\int \frac{d^4k}{(2\pi)^4}\,\sigma^{\alpha\beta\lambda}(k)\,
G_{\beta\lambda}(k)\,e^{\textstyle -ik\cdot x},
\end{eqnarray}
where we have used the relation $f^{abc}f^{dbc} = N_C\,\delta^{ad}$
and denoted $j^{\alpha}(k) =
\sigma^{\alpha\beta\lambda}(k)\,G_{\beta\lambda}(k)$ [3]. The tensor
$\sigma^{\alpha\beta\lambda}(k)$ is the colour conductivity tensor
determined by Eq.(\ref{label2.1}) [3].

The components of the colour conductivity tensor required for the
calculation of the chromoelectric permeability tensor are determined
by
\begin{eqnarray*}
\hspace{-0.3in}&&\sigma^{i 0 j}(k) =
2i\,g^2_s\,\frac{\gamma}{T}\int\frac{d^3p}{(2\pi)^3}\,p^i\nonumber\\
\hspace{-0.3in}&&\times\,\Bigg\{\Bigg(\frac{p^j}{|\vec{p}\,|} -
v^j\Bigg)\Bigg[\frac{N_C}{(\omega + i\nu_g\gamma)\,|\vec{p}\,| -
(\vec{k} + i\nu_g\gamma\vec{v})\cdot \vec{p}}\,\frac{\displaystyle
e^{\textstyle \gamma\,(|\vec{p}\,| - \vec{v}\cdot
\vec{p}\,)/T}}{\displaystyle \Big(e^{\textstyle \gamma\,(|\vec{p}\,| -
\vec{v}\cdot \vec{p}\,)/T} - 1\Big)^{\!2}} \nonumber\\
\hspace{-0.3in}&& + \frac{1}{(\omega + i\nu_q\gamma)\,|\vec{p}\,| -
(\vec{k} + i\nu_q\gamma\vec{v})\cdot \vec{p}}\,\frac{\displaystyle
e^{\textstyle \gamma\,(|\vec{p}\,| - \vec{v}\cdot \vec{p} -
\mu(T)/\gamma)/T}}{\displaystyle \Big(e^{\textstyle
\gamma\,(|\vec{p}\,| - \vec{v}\cdot \vec{p} - \mu(T)/\gamma)/T} +
1\Big)^{\!2}}\nonumber\\
\hspace{-0.3in}&& + \frac{1}{(\omega +
i\nu_q\gamma)\,|\vec{p}\,| - (\vec{k} +
i\nu_q\gamma\vec{v})\cdot \vec{p}}\,\frac{\displaystyle
e^{\textstyle \gamma\,(|\vec{p}\,| - \vec{v}\cdot \vec{p} +
\mu(T)/\gamma)/T}}{\displaystyle \Big(e^{\textstyle
\gamma\,(|\vec{p}\,| - \vec{v}\cdot \vec{p} + \mu(T)/\gamma)/T} +
1\Big)^{\!2}}\Bigg] \nonumber\\
\hspace{-0.3in}&&+ \frac{1}{(\omega +
i\nu_s\gamma)\,\sqrt{\vec{p}^{\,\,2} + m^2_s} - (\vec{k} +
i\nu_s\gamma\vec{v})\cdot
\vec{p}}\,\Bigg(\frac{p^j}{\sqrt{\vec{p}^{\,\,2} + m^2_s}} -
v^j\Bigg)\nonumber\\
\hspace{-0.3in}&&\times\,\frac{\displaystyle e^{\textstyle \gamma\,
(\sqrt{\vec{p}^{\,\,2} + m^2_s} - \vec{v}\cdot \vec{p}\,)/T
}}{\displaystyle \Big(e^{\textstyle \gamma\, (\sqrt{\vec{p}^{\,\,2} +
m^2_s} - \vec{v}\cdot \vec{p}\,)/T } +
1\Big)^{\!2}}\Bigg\},\nonumber\\
\hspace{-0.3in}&&\sigma^{i \ell j}(k)
=2i\,g^2_s\,\frac{\gamma}{T}
\int\frac{d^3p}{(2\pi)^3}\,p^i\,p^{\ell}\nonumber\\
\hspace{-0.3in}&&\times\,\Bigg\{\Bigg(\frac{p^j}{|\vec{p}\,|} -
v^j\Bigg)\,\frac{1}{|\vec{p}\,|}\,
\Bigg[\frac{N_C}{(\omega + i\nu_g\gamma)\,|\vec{p}\,| -
(\vec{k} + i\nu_g\gamma\vec{v})\cdot \vec{p}}\,\frac{\displaystyle
e^{\textstyle \gamma\,(|\vec{p}\,| - \vec{v}\cdot
\vec{p}\,)/T}}{\displaystyle \Big(e^{\textstyle \gamma\,(|\vec{p}\,| -
\vec{v}\cdot \vec{p}\,)/T} - 1\Big)^{\!2}}
\end{eqnarray*}
\begin{eqnarray}\label{label2.10} 
\hspace{-0.3in}&& + \frac{1}{(\omega + i\nu_q\gamma)\,|\vec{p}\,| -
(\vec{k} + i\nu_q\gamma\vec{v})\cdot \vec{p}}\,\frac{\displaystyle
e^{\textstyle \gamma\,(|\vec{p}\,| - \vec{v}\cdot \vec{p} -
\mu(T)/\gamma)/T}}{\displaystyle \Big(e^{\textstyle
\gamma\,(|\vec{p}\,| - \vec{v}\cdot \vec{p} - \mu(T)/\gamma)/T} +
1\Big)^{\!2}}\nonumber\\
\hspace{-0.3in}&& + \frac{1}{(\omega +
i\nu_q\gamma)\,|\vec{p}\,| - (\vec{k} +
i\nu_q\gamma\vec{v})\cdot \vec{p}}\,\frac{\displaystyle
e^{\textstyle \gamma\,(|\vec{p}\,| - \vec{v}\cdot \vec{p} +
\mu(T)/\gamma)/T}}{\displaystyle \Big(e^{\textstyle
\gamma\,(|\vec{p}\,| - \vec{v}\cdot \vec{p} + \mu(T)/\gamma)/T} +
1\Big)^{\!2}}\Bigg] \nonumber\\
\hspace{-0.3in}&&+ \frac{1}{(\omega +
i\nu_s\gamma)\,\sqrt{\vec{p}^{\,\,2} + m^2_s} - (\vec{k} +
i\nu_s\gamma\vec{v})\cdot
\vec{p}}\,\Bigg(\frac{p^j}{\sqrt{\vec{p}^{\,\,2} + m^2_s}} -
v^j\Bigg)\nonumber\\
\hspace{-0.3in}&&\times\,\frac{1}{\sqrt{\vec{p}^{\,\,2} +
m^2_s}}\,\frac{\displaystyle e^{\textstyle \gamma\,
(\sqrt{\vec{p}^{\,\,2} + m^2_s} - \vec{v}\cdot \vec{p}\,)/T
}}{\displaystyle \Big(e^{\textstyle \gamma\, (\sqrt{\vec{p}^{\,\,2} +
m^2_s} - \vec{v}\cdot \vec{p}\,)/T } + 1\Big)^{\!2}}\Bigg\},
\end{eqnarray}
where we have set $\nu_u = \nu_d = \nu_{\bar{u}} = \nu_{\bar{d}} =
\nu_q$ and $\nu_{\bar{s}} = \nu_s$. This is justified by $C$, $P$ and
$T$ invariance of strong interactions.

Now we can proceed to calculating the components of the colour
conductivity tensor Eq.(\ref{label2.10}) for the QGP in different
coordinate frames.

\section{Quark--gluon plasma at rest}
\setcounter{equation}{0}

\hspace{0.2in} The components of the colour conductivity tensor for
the QGP at rest ($\vec{v} = 0$ and $\gamma = 1$) and $\mu(T) = m_s =
0$ have been calculated by Mr${\acute{\rm o}}$wczy${\acute{\rm n}}$ski
[3]. Our results should differ from his calculations by non--zero
values of a chemical potential of light quarks $\mu(T)$ and the
$s$--quark mass $m_s$. Setting $\vec{v} = 0$ and $\gamma = 1$
the components of the colour conductivity tensor given by
Eq.(\ref{label2.10}) acquire the form
\begin{eqnarray}\label{label3.1}
\hspace{-0.3in}&&\sigma^{i 0 j}(k) =
\frac{2ig^2_s}{T}\int\frac{d^3p}{(2\pi)^3}\frac{p^i
p^j}{|\vec{p}\,|}\Bigg\{\Bigg[\frac{N_C}{(\omega +
i\nu_g)|\vec{p}\,| - \vec{k}\cdot \vec{p}}\frac{\displaystyle
e^{\textstyle |\vec{p}\,|/T}}{\displaystyle \Big(e^{\textstyle
|\vec{p}\,|/T} - 1\Big)^{\!2}} + \frac{1}{(\omega +
i\nu_q)|\vec{p}\,| - \vec{k}\cdot \vec{p}}\nonumber\\
\hspace{-0.3in}&&\times\frac{\displaystyle e^{\textstyle
(|\vec{p}\,|- \mu(T))/T}}{\displaystyle \Big(e^{\textstyle
(|\vec{p}\,| - \mu(T))/T} + 1\Big)^{\!2}} + \frac{1}{(\omega +
i\nu_q)|\vec{p}\,| - \vec{k}\cdot
\vec{p}}\frac{\displaystyle e^{\textstyle (|\vec{p}\,| +
\mu(T))/T}}{\displaystyle \Big(e^{\textstyle (|\vec{p}\,| + \mu(T))/T}
+ 1\Big)^{\!2}}\Bigg] \nonumber\\
\hspace{-0.3in}&& + \frac{|\vec{p}\,|}{\sqrt{\vec{p}^{\,\,2} +
m^2_s}}\,\frac{1}{(\omega + i\nu_s)\sqrt{\vec{p}^{\,\,2} +
m^2_s} - \vec{k}\cdot \vec{p}}\frac{\displaystyle e^{\textstyle
\sqrt{\vec{p}^{\,\,2} + m^2_s}/T }}{\displaystyle \Big(e^{\textstyle
\sqrt{\vec{p}^{\,\,2} + m^2_s}/T} + 1\Big)^{\!2}}\Bigg\},\nonumber\\
\hspace{-0.3in}&&\sigma^{i \ell j}(k)
=\frac{2ig^2_s}{T}\int\frac{d^3p}{(2\pi)^3}\frac{p^i p^{\ell}
p^j}{\vec{p}^{\,\,2}}\Bigg\{ \Bigg[\frac{N_C}{(\omega +
i\nu_g)|\vec{p}\,| - \vec{k}\cdot \vec{p}}\frac{\displaystyle
e^{\textstyle |\vec{p}\,|/T}}{\displaystyle \Big(e^{\textstyle
|\vec{p}\,|/T} - 1\Big)^{\!2}} + \frac{1}{(\omega +
i\nu_q)|\vec{p}\,| - \vec{k} \cdot \vec{p}}\nonumber\\
\hspace{-0.3in}&&\times\,\frac{\displaystyle e^{\textstyle
(|\vec{p}\,| - \mu(T))/T}}{\displaystyle \Big(e^{\textstyle
(|\vec{p}\,| - \mu(T))/T} + 1\Big)^{\!2}} + \frac{1}{(\omega +
i\nu_q)|\vec{p}\,| - \vec{k}\cdot \vec{p}}\frac{\displaystyle
e^{\textstyle (|\vec{p}\,| + \mu(T))/T}}{\displaystyle
\Big(e^{\textstyle (|\vec{p}\,| + \mu(T))/T} + 1\Big)^{\!2}}\Bigg]
\nonumber\\
\hspace{-0.3in}&&+ \frac{\vec{p}^{\,\,2}}{\vec{p}^{\,\,2} +
m^2_s}\,\frac{1}{(\omega + i\nu_s)\sqrt{\vec{p}^{\,\,2} + m^2_s} -
\vec{k}\cdot \vec{p}}\frac{\displaystyle e^{\textstyle
\sqrt{\vec{p}^{\,\,2} + m^2_s}/T }}{\displaystyle \Big(e^{\textstyle
\sqrt{\vec{p}^{\,\,2} + m^2_s}/T } + 1\Big)^{\!2}}\Bigg\}.
\end{eqnarray}
The components $\sigma^{i \ell j}(k)$ enter to the definition of the
chromoelectric permeability determined by Eq.(\ref{label1.14}) in the
antisymmetric combination $(\sigma^{i j \ell}(k) - \sigma^{i \ell
j}(k))$.  Since the tensor $\sigma^{i\ell j}$ defined by
Eq.(\ref{label3.1}) is fully symmetric, so it is obvious that
$(\sigma^{i j \ell}(k) - \sigma^{i \ell j}(k)) = 0$. Thereby, a
non--trivial contribution of the colour conductivity tensor to the
chromoelectric permeability of the QGP at rest comes from the
components $\sigma^{i 0 j}(k)$ given by Eq.(\ref{label3.1}).

Since the tensor $\sigma^{i 0 j}(k)$ is symmetric, $\sigma^{i 0 j}(k) =
\sigma^{j 0 i}(k)$, the general expression of $\sigma^{i 0 j}(k)$ can
written as
\begin{eqnarray}\label{label3.2}
\sigma^{i 0 j}(k) = i\,A(\omega, \vec{k}\,)\,\Big(\delta^{i j} -
\frac{k^i k^j}{\vec{k}^{\,2}}\Big) + i\,B(\omega, \vec{k}\,)\,
\frac{k^i k^j}{\vec{k}^{\,2}}.
\end{eqnarray}
The coefficient functions $A(\omega, \vec{k}\,)$ and $B(\omega,
\vec{k}\,)$ are defined by the integrals
\begin{eqnarray}\label{label3.3}
\hspace{-0.3in}&&A(\omega, \vec{k}\,) = - \frac{1}{2}\,B(\omega,
\vec{k}\,)\nonumber\\
\hspace{-0.3in}&& + \frac{g^2_s}{T}
\int\frac{d^3p}{(2\pi)^3}\, \Bigg\{\Bigg[\frac{N_C}{(\omega + i\nu_g) -
\vec{k}\cdot \vec{n}}\,\frac{\displaystyle e^{\textstyle
|\vec{p}\,|/T}}{\displaystyle \Big(e^{\textstyle |\vec{p}\,|/T} -
1\Big)^{\!2}} + \frac{1}{(\omega + i\nu_q) - \vec{k}\cdot
\vec{n}}\nonumber\\
\hspace{-0.3in}&&\times\,\frac{\displaystyle e^{\textstyle
(|\vec{p}\,|- \mu(T))/T}}{\displaystyle \Big(e^{\textstyle
(|\vec{p}\,| - \mu(T))/T} + 1\Big)^{\!2}} + \frac{1}{(\omega +
i\nu_q) - \vec{k}\cdot \vec{n}}\,\frac{\displaystyle
e^{\textstyle (|\vec{p}\,| + \mu(T))/T}}{\displaystyle
\Big(e^{\textstyle (|\vec{p}\,| + \mu(T))/T} + 1\Big)^{\!2}}\,\Bigg]
\nonumber\\
\hspace{-0.3in}&& + \frac{\vec{p}^{\,\,2}}{\sqrt{\vec{p}^{\,\,2} +
m^2_s}}\,\frac{1}{(\omega + i\nu_s)\,\sqrt{\vec{p}^{\,\,2} + m^2_s} -
\vec{k}\cdot \vec{p}}\,\frac{\displaystyle e^{\textstyle
\sqrt{\vec{p}^{\,\,2} + m^2_s}/T }}{\displaystyle \Big(e^{\textstyle
\sqrt{\vec{p}^{\,\,2} + m^2_s}/T} + 1\Big)^{\!2}}\Bigg\},\nonumber\\
\hspace{-0.3in}&&B(\omega, \vec{k}\,) =
\frac{2g^2_s}{T}\,\frac{1}{\vec{k}^{\,2}} \int\frac{d^3p}{(2\pi)^3}\,
\Bigg\{\Bigg[N_C\,\frac{(\vec{n}\cdot \vec{k}\,)^2}{(\omega + \nu_g) -
\vec{k}\cdot \vec{n}}\,\frac{\displaystyle e^{\textstyle
|\vec{p}\,|/T}}{\displaystyle \Big(e^{\textstyle |\vec{p}\,|/T} -
1\Big)^{\!2}} + \frac{(\vec{n}\cdot \vec{k}\,)^2}{(\omega + \nu_q) -
\vec{k}\cdot \vec{n}}\nonumber\\
\hspace{-0.3in}&&\times\,\frac{\displaystyle e^{\textstyle
(|\vec{p}\,|- \mu(T))/T}}{\displaystyle \Big(e^{\textstyle
(|\vec{p}\,| - \mu(T))/T} + 1\Big)^{\!2}} + \frac{(\vec{n}\cdot
\vec{k}\,)^2}{(\omega + i\nu_q) - \vec{k}\cdot
\vec{n}}\,\frac{\displaystyle e^{\textstyle (|\vec{p}\,| +
\mu(T))/T}}{\displaystyle \Big(e^{\textstyle (|\vec{p}\,| + \mu(T))/T}
+ 1\Big)^{\!2}}\,\Bigg] \nonumber\\
\hspace{-0.3in}&& + \frac{\vec{p}^{\,\,2}}{\sqrt{\vec{p}^{\,\,2} +
m^2_s}}\,\frac{(\vec{n}\cdot \vec{k}\,)^2}{(\omega +
i\nu_s)\,\sqrt{\vec{p}^{\,\,2} + m^2_s} - \vec{k}\cdot
\vec{p}}\,\frac{\displaystyle e^{\textstyle \sqrt{\vec{p}^{\,\,2} +
m^2_s}/T }}{\displaystyle \Big(e^{\textstyle \sqrt{\vec{p}^{\,\,2} +
m^2_s}/T} + 1\Big)^{\!2}}\Bigg\},
\end{eqnarray}
where $\vec{n} = \vec{p}/|\vec{p}\,|$.

In terms of the coefficient functions $A(\omega, \vec{k}\,)$ and
$B(\omega, \vec{k}\,)$ the transversal
$\varepsilon_T(\omega,\vec{k}\,)$ and longitudinal
$\varepsilon_L(\omega,\vec{k}\,)$ parts of the chromoelectric
permeability tensor $\varepsilon^{ij}(\omega,\vec{k}\,)$ are given by
\begin{eqnarray}\label{label3.4}
\varepsilon_T(\omega,\vec{k}\,) &=& 1 + \frac{1}{\omega}\,A(\omega,
\vec{k}\,),\nonumber\\ \varepsilon_L(\omega,\vec{k}\,) &=& 1 +
\frac{1}{\omega}\,B(\omega, \vec{k}\,) .
\end{eqnarray}
A direct calculation of the integrals over $\vec{p}$ in
Eq.(\ref{label3.3}) yields
\begin{eqnarray*}
\hspace{-0.5in}&&A(\omega, \vec{k}\,) = \frac{g^2_sN_C}{12}\,
\frac{T^2}{|\vec{k}\,|}\,\Bigg\{\Bigg[2\,\frac{\omega +
i\nu_g}{|\vec{k}\,|} + \Bigg(1 - \frac{(\omega +
i\nu_g)^2}{\vec{k}^{\,2}}\Bigg)\,{\ell n}\Bigg(\frac{\omega + i\nu_g +
|\vec{k}\,|}{\omega + i\nu_g - |\vec{k}\,|}\Bigg)\Bigg]\nonumber\\
\hspace{-0.5in}&& + \frac{1}{N_C}\,\Bigg(1 +
\frac{3}{\pi^2}\,\frac{\mu^2(T)}{T^2}\Bigg)\Bigg[2\,\frac{\omega +
i\nu_q}{|\vec{k}\,|} + \Bigg(1 - \frac{(\omega +
i\nu_q)^2}{\vec{k}^{\,2}}\Bigg)\,{\ell n}\Bigg(\frac{\omega + i\nu_q +
|\vec{k}\,|}{\omega + i\nu_q - |\vec{k}\,|}\Bigg)\Bigg]\nonumber\\
\hspace{-0.5in}&& +
\frac{1}{N_C}\,\frac{6}{\pi^2}\,\frac{1}{T^2}\int^{\infty}_{0}
\frac{dp\,p}{\displaystyle e^{\textstyle \sqrt{p^2 + m^2_s}/T}
+1}\Bigg[{\ell n}\Bigg(\frac{(\omega + i\nu_s)\sqrt{p^2 + m^2_s} +
|\vec{k}\,|p}{(\omega + i\nu_s)\sqrt{p^2 + m^2_s} -
|\vec{k}\,|p}\Bigg)
\end{eqnarray*}
\begin{eqnarray*}
\hspace{-0.5in}&& + \frac{|\vec{k}\,|(\omega + i\nu_s)m^2_s}{(\omega +
i\nu_s)^2(p^2 + m^2_s) - \vec{k}^{\,2}p^2}\,\frac{\sqrt{p^2 +
m^2_s}}{p}\Bigg]\nonumber\\
\hspace{-0.5in}&& +
\frac{1}{N_C}\,\frac{6}{\pi^2}\,\frac{1}{T^2}\int^{\infty}_0
\frac{dp\,p}{\displaystyle e^{\textstyle \sqrt{p^2 + m^2_s}/T} +
1}\,\Bigg[2\,\frac{\omega + i\nu_s}{|\vec{k}\,|}\frac{p^2 +
m^2_s/2}{p\sqrt{p^2 + m^2_s}} - \frac{(\omega +
i\nu_s)^2}{\vec{k}^{\,2}} \nonumber\\
\hspace{-0.5in}&&\times\,\Bigg[{\ell n}\Bigg(\frac{(\omega +
i\nu_s)\sqrt{p^2 + m^2_s} + |\vec{k}\,|p}{(\omega + i\nu_s)\sqrt{p^2 +
m^2_s} - |\vec{k}\,|p}\Bigg) + \frac{|\vec{k}\,|(\omega +
i\nu_s)m^2_s}{(\omega + i\nu_s)^2(p^2 + m^2_s) -
\vec{k}^{\,2}p^2}\,\frac{\sqrt{p^2 + m^2_s}}{p}\Bigg]\Bigg\},
\end{eqnarray*}
\begin{eqnarray}\label{label3.5}
\hspace{-0.5in}&&B(\omega, \vec{k}\,) = -\frac{g^2_sN_C}{6}\,
\frac{T^2}{|\vec{k}\,|}\,\Bigg\{\Bigg[2\,\frac{\omega +
i\nu_g}{|\vec{k}\,|} - \frac{(\omega + i\nu_g)^2}{\vec{k}^{\,2}}\,{\ell
n}\Bigg(\frac{\omega + i\nu_g + |\vec{k}\,|}{\omega + i\nu_g -
|\vec{k}\,|}\Bigg)\Bigg]\nonumber\\
\hspace{-0.5in}&& + \frac{1}{N_C}\,\Bigg(1 +
\frac{3}{\pi^2}\,\frac{\mu^2(T)}{T^2}\Bigg)\Bigg[2\,\frac{\omega +
i\nu_q}{|\vec{k}\,|} - \frac{(\omega + i\nu_q)^2}{\vec{k}^{\,2}}\,{\ell
n}\Bigg(\frac{\omega + i\nu_q + |\vec{k}\,|}{\omega + i\nu_q -
|\vec{k}\,|}\Bigg)\Bigg]\nonumber\\
\hspace{-0.5in}&& +
\frac{1}{N_C}\,\frac{3}{\pi^2}\,\frac{1}{T^2}\int^{\infty}_0
\frac{dp}{\displaystyle e^{\textstyle \sqrt{p^2 + m^2_s}/T} +
1}\frac{d}{dp}\Bigg[2\,\frac{\omega + i\nu_s}{|\vec{k}\,|}p\sqrt{p^2 +
m^2_s} - \frac{(\omega + i\nu_s)^2}{\vec{k}^{\,2}}(p^2 +
m^2_s)\nonumber\\
\hspace{-0.5in}&&\times\,{\ell n}\Bigg(\frac{(\omega + i\nu_s)\sqrt{p^2
+ m^2_s} + |\vec{k}\,|p}{(\omega + i\nu_s)\sqrt{p^2 + m^2_s} -
|\vec{k}\,|p}\Bigg)\Bigg]\Bigg\} = \nonumber\\
\hspace{-0.5in}&&= - \frac{g^2_sN_C}{6}\,
\frac{T^2}{|\vec{k}\,|}\,\Bigg\{\Bigg[2\,\frac{\omega +
i\nu_g}{|\vec{k}\,|} - \frac{(\omega +
i\nu_g)^2}{\vec{k}^{\,2}}\,{\ell n}\Bigg(\frac{\omega + i\nu_g +
|\vec{k}\,|}{\omega + i\nu_g - |\vec{k}\,|}\Bigg)\Bigg]\nonumber\\
\hspace{-0.5in}&& + \frac{1}{N_C}\,\Bigg(1 +
\frac{3}{\pi^2}\,\frac{\mu^2(T)}{T^2}\Bigg)\Bigg[2\,\frac{\omega +
i\nu_q}{|\vec{k}\,|} - \frac{(\omega + i\nu_q)^2}{\vec{k}^{\,2}}\,{\ell
n}\Bigg(\frac{\omega + i\nu_q + |\vec{k}\,|}{\omega + i\nu_q -
|\vec{k}\,|}\Bigg)\Bigg]\nonumber\\
\hspace{-0.5in}&& +
\frac{1}{N_C}\,\frac{6}{\pi^2}\,\frac{1}{T^2}\int^{\infty}_0
\frac{dp\,p}{\displaystyle e^{\textstyle \sqrt{p^2 + m^2_s}/T} +
1}\,\Bigg[2\,\frac{\omega + i\nu_s}{|\vec{k}\,|}\frac{p^2 +
m^2_s/2}{p\sqrt{p^2 + m^2_s}} - \frac{(\omega +
i\nu_s)^2}{\vec{k}^{\,2}} \nonumber\\
\hspace{-0.5in}&&\times\,\Bigg[{\ell n}\Bigg(\frac{(\omega +
i\nu_s)\sqrt{p^2 + m^2_s} + |\vec{k}\,|p}{(\omega + i\nu_s)\sqrt{p^2 +
m^2_s} - |\vec{k}\,|p}\Bigg) + \frac{|\vec{k}\,|(\omega +
i\nu_s)m^2_s}{(\omega + i\nu_s)^2(p^2 + m^2_s) -
\vec{k}^{\,2}p^2}\,\frac{\sqrt{p^2 + m^2_s}}{p}\Bigg]\Bigg\}.
\end{eqnarray}
For a numerical estimate of the contribution of a chemical potential
to the chromoelectric permeability tensor we would use the chemical
potential calculated in [14]. A temperature we set equal to a
freeze--out temperature $T = T_f$, a typical value of which amounts to
$T_f = 175\,{\rm MeV}$ for ultra--relativistic heavy--ion
collisions. At $T = T_f = 175\,{\rm MeV}$ one can find that $\mu(T)/T
\simeq 0.29$. This means that the contribution of a chemical potential
of light quarks is insignificant for the formation of the
chromoelectric permeability of the QGP, and without loss of generality
one can set $\mu(T) = 0$.

For the estimate of the contribution of the non--zero value of the
$s$--quark mass we would notice that the main contribution to the
integral over $p$ comes from the region $p \le T$. Due to this the
$\omega$ and $|\vec{k}\,|$ dependence of the chromoelectric
permeability induced by strange quarks and antiquarks relative to the
contribution of gluons and light quarks and antiquarks enters with the
factor
\begin{eqnarray}\label{label3.6}
\frac{1}{N_C + 1}\,\frac{6}{\pi^2}\,\Big[F\Big(e^{\textstyle
-m_s/T}\Big) + \frac{m_s}{T}\,{\ell n}\Big(1 + e^{\textstyle
-m_s/T}\Big)\Big],
\end{eqnarray}
where $F(x)$ is Spence's function [15]
\begin{eqnarray}\label{label3.7}
F(x) = \int^1_0\frac{dt}{t}\,{\ell n}(1 + xt).
\end{eqnarray}
At $N_C = 3$, $m_s = 135\,{\rm MeV}$ and $T = 175\,{\rm MeV}$ the
factor Eq.(\ref{label3.6}) is less than $0.11$. This implies the
possibility to neglect the contribution of massive strange quarks and
antiquarks to the chromoelectric permeability at the freeze--out
temperature.

Thus, according to our estimate the coefficient functions
Eq.(\ref{label3.5}) read now
\begin{eqnarray}\label{label3.8}
\hspace{-0.5in}&&A(\omega, \vec{k}\,) = \frac{g^2_sN_C}{6}\,
\frac{T^2}{|\vec{k}\,|}\,\Bigg\{\Bigg[\frac{\omega +
i\nu_g}{|\vec{k}\,|} + \frac{1}{2}\,\Bigg(1 - \frac{(\omega +
i\nu_g)^2}{\vec{k}^{\,2}}\Bigg)\,{\ell n}\Bigg(\frac{\omega + i\nu_g +
|\vec{k}\,|}{\omega + i\nu_g - |\vec{k}\,|}\Bigg)\Bigg]\nonumber\\
\hspace{-0.5in}&& + \frac{1}{N_C}\,\Bigg[\frac{\omega +
i\nu_q}{|\vec{k}\,|} + \frac{1}{2}\,\Bigg(1 - \frac{(\omega +
i\nu_q)^2}{\vec{k}^{\,2}}\Bigg)\,{\ell n}\Bigg(\frac{\omega + i\nu_q +
|\vec{k}\,|}{\omega + i\nu_q -
|\vec{k}\,|}\Bigg)\Bigg]\Bigg\},\nonumber\\
\hspace{-0.5in}&&B(\omega, \vec{k}\,) = - \frac{g^2_sN_C}{3}\,
\frac{T^2}{|\vec{k}\,|}\,\Bigg\{\Bigg[\frac{\omega +
i\nu_g}{|\vec{k}\,|} - \frac{1}{2}\,\frac{(\omega +
i\nu_g)^2}{\vec{k}^{\,2}}\,{\ell n}\Bigg(\frac{\omega + i\nu_g +
|\vec{k}\,|}{\omega + i\nu_g - |\vec{k}\,|}\Bigg)\Bigg]\nonumber\\
\hspace{-0.5in}&& + \frac{1}{N_C}\,\Bigg[\frac{\omega +
\nu_q}{|\vec{k}\,|} - \frac{1}{2}\,\frac{(\omega +
i\nu_q)^2}{\vec{k}^{\,2}}\,{\ell n}\Bigg(\frac{\omega + i\nu_q +
|\vec{k}\,|}{\omega + i\nu_q - |\vec{k}\,|}\Bigg)\Bigg]\Bigg\}.
\end{eqnarray}
In order to compare our results with those calculated in [2,3] we
should set $\nu_g = \nu_q = \nu$. This gives
\begin{eqnarray}\label{label3.9}
\hspace{-0.5in}&&A(\omega, \vec{k}\,) = \frac{g^2_s(N_C + 1)}{6}\,
\frac{T^2}{|\vec{k}\,|}\,\Bigg[\frac{\omega + i\nu}{|\vec{k}\,|} +
\frac{1}{2}\,\Bigg(1 - \frac{(\omega +
i\nu)^2}{\vec{k}^{\,2}}\Bigg)\,{\ell n}\Bigg(\frac{\omega + i\nu +
|\vec{k}\,|}{\omega + i\nu - |\vec{k}\,|}\Bigg)\Bigg],\nonumber\\
\hspace{-0.5in}&&B(\omega, \vec{k}\,) = - \frac{g^2_s(N_C + 1)}{3}\,
\frac{T^2}{|\vec{k}\,|}\,\Bigg[\frac{\omega +
i\nu}{|\vec{k}\,|} - \frac{1}{2}\,\frac{(\omega +
i\nu)^2}{\vec{k}^{\,2}}\,{\ell n}\Bigg(\frac{\omega + i\nu +
|\vec{k}\,|}{\omega + i\nu - |\vec{k}\,|}\Bigg)\Bigg].
\end{eqnarray}
The transversal and longitudinal components of the chromoelectric
permeability tensor we obtain in the form
\begin{eqnarray}\label{label3.10}
\varepsilon_T(\omega,\vec{k}\,) &=& 1 + \frac{g^2_s(N_C +
1)}{12\,\omega}\, \frac{T^2}{|\vec{k}\,|}\,\Bigg[\frac{\omega +
i\nu}{|\vec{k}\,|} + \frac{1}{2}\,\Bigg(1 - \frac{(\omega +
i\nu)^2}{\vec{k}^{\,2}}\Bigg)\,{\ell n}\Bigg(\frac{\omega + i\nu +
|\vec{k}\,|}{\omega + i\nu - |\vec{k}\,|}\Bigg)\Bigg],\nonumber\\
\varepsilon_L(\omega,\vec{k}\,) &=& 1 - \frac{g^2_s(N_C +
1)}{6\,\omega}\, \frac{T^2}{|\vec{k}\,|}\,\Bigg[\frac{\omega +
i\nu}{|\vec{k}\,|} - \frac{1}{2}\,\frac{(\omega +
i\nu)^2}{\vec{k}^{\,2}}\,{\ell n}\Bigg(\frac{\omega + i\nu +
|\vec{k}\,|}{\omega + i\nu - |\vec{k}\,|}\Bigg)\Bigg].
\end{eqnarray}
Following [2,3] and introducing the plasma frequency 
\begin{eqnarray}\label{label3.11}
\omega^2_0 = \frac{1}{3}\,g^2_s\,(N_C + 1)\,T^2
\end{eqnarray}
we obtain
\begin{eqnarray}\label{label3.12}
\varepsilon_T(\omega,\vec{k}\,) &=& 1 + \frac{1}{\omega}\,
\frac{\omega^2_0}{2|\vec{k}\,|}\, \Bigg[\frac{\omega +
\nu}{|\vec{k}\,|} + \frac{1}{2}\,\Bigg(1 - \frac{(\omega +
i\nu)^2}{\vec{k}^{\,2}}\Bigg)\,{\ell n}\Bigg(\frac{\omega + i\nu +
|\vec{k}\,|}{\omega + i\nu - |\vec{k}\,|}\Bigg)\Bigg],\nonumber\\
\varepsilon_L(\omega,\vec{k}\,) &=& 1 - \frac{1}{\omega}\,
\frac{\omega^2_0}{|\vec{k}\,|}\,\Bigg[\frac{\omega + i\nu}{|\vec{k}\,|}
- \frac{1}{2}\,\frac{(\omega + i\nu)^2}{\vec{k}^{\,2}}\,{\ell
n}\Bigg(\frac{\omega + i\nu + |\vec{k}\,|}{\omega + i\nu -
|\vec{k}\,|}\Bigg)\Bigg].
\end{eqnarray}
These expressions agree well with those calculated in [2,3]. The
plasma frequency (\ref{label3.11}) is diminished with respect to the
expression given in [3]: $\omega^2_0 = g^2_sT^2(2N_C + 3)/6$.  The
former is due to the neglect of the contribution of massive strange
quarks and antiquarks the validity of which has been demonstrated
above.

\section{Ultra--relativistic and  spherical symmetric expanding  QGP}
\setcounter{equation}{0}
 
\hspace{0.2in} According to our estimates carried out in the preceding
section, for the analysis of the chromoelectric permeability of a
moving QGP we would use the chromoelectric conductivity tensor with
components
\begin{eqnarray}\label{label4.1}
\hspace{-0.3in}&&\sigma^{i 0 j}(k) =
2i\,g^2_s\,\frac{\gamma}{T}\int\frac{d^3p}{(2\pi)^3}\,p^i\,
\Bigg(\frac{p^j}{|\vec{p}\,|}
- v^j\Bigg)\frac{1}{(\omega + i\nu\gamma)\,|\vec{p}\,| - (\vec{k} +
i\nu\gamma\vec{v})\cdot \vec{p}}\nonumber\\
\hspace{-0.3in}&&\times\,\Bigg[N_C\,\frac{\displaystyle e^{\textstyle
\gamma\,(|\vec{p}\,| - \vec{v}\cdot \vec{p}\,)/T}}{\displaystyle
\Big(e^{\textstyle \gamma\,(|\vec{p}\,| - \vec{v}\cdot \vec{p}\,)/T} -
1\Big)^{\!2}} + 2\,\frac{\displaystyle e^{\textstyle
\gamma\,(|\vec{p}\,| - \vec{v}\cdot \vec{p})/T}}{\displaystyle
\Big(e^{\textstyle \gamma\,(|\vec{p}\,| - \vec{v}\cdot \vec{p})/T} +
1\Big)^{\!2}}\Bigg],\nonumber\\
\hspace{-0.3in}&&\sigma^{i \ell j}(k) = 2i\,g^2_s\,\frac{\gamma}{T}
\int\frac{d^3p}{(2\pi)^3}\,p^i\,p^{\ell}\Bigg(\frac{p^j}{|\vec{p}\,|}
- v^j\Bigg)\,\frac{1}{|\vec{p}\,|}\,\frac{1}{(\omega +
i\nu\gamma)\,|\vec{p}\,| - (\vec{k} + i\nu\gamma\vec{v})\cdot \vec{p}}
\nonumber\\
\hspace{-0.3in}&&\times\,\Bigg[N_C\,\frac{\displaystyle e^{\textstyle
\gamma\,(|\vec{p}\,| - \vec{v}\cdot \vec{p}\,)/T}}{\displaystyle
\Big(e^{\textstyle \gamma\,(|\vec{p}\,| - \vec{v}\cdot \vec{p}\,)/T} -
1\Big)^{\!2}} + 2\,\frac{\displaystyle e^{\textstyle
\gamma\,(|\vec{p}\,| - \vec{v}\cdot \vec{p})/T}} {\displaystyle
\Big(e^{\textstyle \gamma\,(|\vec{p}\,| - \vec{v}\cdot \vec{p} )/T} +
1\Big)^{\!2}}\Bigg].
\end{eqnarray}
For ultra--relativistic heavy--ion collisions in the center of mass
frame one can consider the case of an ultra--relativistic and
spherical symmetric expanding QGP.

For the spherical symmetric expanding QGP a hydrodynamical velocity
$\vec{v}$ should be directed along a momentum $\vec{p}$, i.e. $\vec{v}
= v\,\vec{n}$, where $\vec{n} = \vec{p}/|\vec{p}\,|$. In this case the
components of the colour conductivity tensor Eq.(\ref{label4.1}) read
\begin{eqnarray}\label{label4.2}
\hspace{-0.5in}&&\sigma^{i 0 j}(k) =\nonumber\\
\hspace{-0.5in}&&=\frac{2ig^2_s}{T\bar{\gamma}}
\int\frac{d^3p}{(2\pi)^3}\,\frac{p^ip^j}{|\vec{p}\,|}\,
\frac{1}{\displaystyle \Big(\omega +
\frac{i\nu}{\bar{\gamma}}\Big)\,|\vec{p}\,| - \vec{k}\cdot
\vec{p}}\,\Bigg[\frac{\displaystyle N_C\,e^{\textstyle
|\vec{p}\,|/T\bar{\gamma}}}{\displaystyle \Big(e^{\textstyle
|\vec{p}\,|/T\bar{\gamma}} - 1\Big)^{\!2}} + \frac{\displaystyle
2\,e^{\textstyle |\vec{p}\,|/T\bar{\gamma}}}{\displaystyle
\Big(e^{\textstyle |\vec{p}\,|/T\bar{\gamma}} +
1\Big)^{\!2}}\Bigg],\nonumber\\
\hspace{-0.5in}&&\sigma^{i \ell j}(k) =\nonumber\\
\hspace{-0.5in}&&= \frac{2ig^2_s}{T\bar{\gamma}}
\int\frac{d^3p}{(2\pi)^3}\,\frac{p^ip^{\ell}p^j}{\vec{p}^{\,2}}\,
\frac{1}{\displaystyle \Big(\omega +
\frac{i\nu}{\bar{\gamma}}\Big)\,|\vec{p}\,| - \vec{k} \cdot \vec{p}}
\,\Bigg[\frac{\displaystyle N_C\,e^{\textstyle
|\vec{p}\,|/T\bar{\gamma}}}{\displaystyle \Big(e^{\textstyle
|\vec{p}\,|/T\bar{\gamma}} - 1\Big)^{\!2}} + \frac{\displaystyle
2\,e^{\textstyle |\vec{p}\,|/T\bar{\gamma}}}{\displaystyle
\Big(e^{\textstyle |\vec{p}\,|/T\bar{\gamma}} + 1\Big)^{\!2}}\Bigg],
\end{eqnarray}
where we have denoted
\begin{eqnarray}\label{label4.3}
\bar{\gamma} = \sqrt{\frac{1 + v}{1 - v}}.
\end{eqnarray}
In the ultra--relativistic limit $v \simeq 1$ one can set
$\bar{\gamma} \simeq 2\gamma$. The tensor $\sigma^{i \ell j}(k)$ is
symmetric with respect to indices $\ell$ and $j$, $\sigma^{i \ell
j}(k) = \sigma^{i j \ell }(k)$, and, therefore, does not contribute to
the chromoelectric permeability tensor $\varepsilon^{ij}(k)$.

The tensor $\sigma^{i0j}(k)$ can be represented by the integral
\begin{eqnarray}\label{label4.4}
\sigma^{i 0 j}(k) =i\,\frac{1}{3}\,(N_C + 1)\,g^2_s\,T\,\bar{\gamma}^2
\int\frac{d\Omega_{\textstyle \vec{n}}}{4\pi}\,
\frac{n^in^j}{\displaystyle \Big(\omega +
\frac{i\nu}{\bar{\gamma}}\Big) - \vec{k}\cdot \vec{n}}.
\end{eqnarray}
The coefficient functions $A(\omega, \vec{k}\,)$ and $B(\omega,
\vec{k}\,)$ read 
\begin{eqnarray}\label{label4.5}
\hspace{-0.3in}&&A(\omega, \vec{k}\,) =
\bar{\gamma}^2\,\frac{g^2_s(N_C + 1)}{6}\,
\frac{T^2}{|\vec{k}\,|}\,\Bigg[\frac{\omega + i\nu/\bar{\gamma}
}{|\vec{k}\,|} + \frac{1}{2}\,\Bigg(1 - \frac{(\omega +
i\nu/\bar{\gamma} )^2}{\vec{k}^{\,2}}\Bigg)\,{\ell n}\Bigg(\frac{\omega
+ |\vec{k}\,| + i\nu/\bar{\gamma}} {\omega - |\vec{k}\,| +
\nu/\bar{\gamma} }\Bigg)\Bigg],\nonumber\\
\hspace{-0.3in}&&B(\omega, \vec{k}\,) = -
\bar{\gamma}^2\,\frac{g^2_s(N_C + 1)}{3}\,
\frac{T^2}{|\vec{k}\,|}\,\Bigg[\frac{\omega + i\nu/\bar{\gamma}
}{|\vec{k}\,|} - \frac{1}{2}\,\frac{(\omega + i\nu/\bar{\gamma}
)^2}{\vec{k}^{\,2}}\,{\ell n}\Bigg(\frac{\omega + |\vec{k}\,| +
i\nu/\bar{\gamma}}{\omega - |\vec{k}\,| + i\nu/\bar{\gamma}
}\Bigg)\Bigg].
\end{eqnarray}
The transversal and longitudinal components of the chromoelectric
permeability tensor we obtain in the form
\begin{eqnarray}\label{label4.6}
&&\varepsilon_T(\omega,\vec{k}\,) =\nonumber\\ &&= 1 +
\bar{\gamma}^2\,\frac{g^2_s(N_C + 1)}{6}\,
\frac{T^2}{|\vec{k}\,|}\,\Bigg[\frac{\omega + i\nu/\bar{\gamma}
}{|\vec{k}\,|} + \frac{1}{2}\,\Bigg(1 - \frac{(\omega +
i\nu/\bar{\gamma} )^2}{\vec{k}^{\,2}}\Bigg)\,{\ell n}\Bigg(\frac{\omega
+ |\vec{k}\,| + i\nu/\bar{\gamma}} {\omega - |\vec{k}\,| +
\nu/\bar{\gamma} }\Bigg)\Bigg],\nonumber\\
&&\varepsilon_L(\omega,\vec{k}\,) =\nonumber\\ && = 1 -
\bar{\gamma}^2\,\frac{g^2_s(N_C + 1)}{3}\,
\frac{T^2}{|\vec{k}\,|}\,\Bigg[\frac{\omega + i\nu/\bar{\gamma}
}{|\vec{k}\,|} - \frac{1}{2}\,\frac{(\omega + i\nu/\bar{\gamma}
)^2}{\vec{k}^{\,2}}\,{\ell n}\Bigg(\frac{\omega + |\vec{k}\,| +
i\nu/\bar{\gamma}}{\omega - |\vec{k}\,| + i\nu/\bar{\gamma}
}\Bigg)\Bigg].  \
\end{eqnarray}
In the ultra--relativistic limit, when $\bar{\gamma} \simeq 2\gamma$
and $\gamma \gg 1$, we can neglect the contribution of the term
$\nu/\bar{\gamma}$ and get
\begin{eqnarray}\label{label4.7}
&&\varepsilon_T(\omega,\vec{k}\,) = 1 +
\bar{\gamma}^2\,\frac{g^2_s(N_C + 1)}{6\,\omega}\,
\frac{T^2}{\vec{k}^{\,2}}\,\Bigg[1 - \frac{1}{2}\,\Bigg(
\frac{\omega}{|\vec{k}\,|} - \frac{|\vec{k}\,|}{ \omega}\Bigg) \,{\ell
n}\Bigg(\frac{\omega + |\vec{k}\,|}{\omega -
|\vec{k}\,|}\Bigg)\Bigg],\nonumber\\ &&\varepsilon_L(\omega,\vec{k}\,)
= 1 - \bar{\gamma}^2\,\frac{g^2_s(N_C + 1)}{3\,\omega}\,
\frac{T^2}{\vec{k}^{\,2}}\,\Bigg[1 - \frac{1}{2}\,\frac{\omega
}{|\vec{k}\,|}\,{\ell n}\Bigg(\frac{\omega + |\vec{k}\,| }{\omega -
|\vec{k}\,|}\Bigg)\Bigg].
\end{eqnarray}
These expressions can be transcribed in terms of the
ultra--relativistic, spherical symmetric expanding plasma frequency
\begin{eqnarray}\label{label4.8}
\bar{\omega}^2_0 = \frac{1}{3}\,\bar{\gamma}^2\,g^2_s\,(N_C +1)\,T^2 =
\bar{\gamma}^2\,\omega^2_0.
\end{eqnarray}
In terms of the plasma frequency $\bar{\omega}^2_0$ the transversal
and longitudinal components of the chromoelectric permeability tensor
read
\begin{eqnarray}\label{label4.9}
&&\varepsilon_T(\omega,\vec{k}\,) = 1 + \frac{1}{2}\,
\frac{\bar{\omega}^2_0}{\vec{k}^{\,2}}\,\Bigg[1 - \frac{1}{2}\,\Bigg(
\frac{\omega}{|\vec{k}\,|} - \frac{|\vec{k}\,|}{ \omega}\Bigg) \,{\ell
n}\Bigg(\frac{\omega + |\vec{k}\,|}{\omega -
|\vec{k}\,|}\Bigg)\Bigg],\nonumber\\ &&\varepsilon_L(\omega,\vec{k}\,)
= 1 - \frac{\bar{\omega}^2_0}{\vec{k}^{\,2}}\,\Bigg[1 -
\frac{1}{2}\,\frac{\omega }{|\vec{k}\,|}\,{\ell n}\Bigg(\frac{\omega +
|\vec{k}\,| }{\omega - |\vec{k}\,|}\Bigg)\Bigg].
\end{eqnarray}
These expressions are in analytical agreement with those obtained in
Refs.[2,3] for the QGP at rest.

It is obvious that the expressions Eq.(\ref{label4.9}) for the
transversal and longitudinal components of the chromoelectric
permeability tensor should be retained even if the inverse relaxation
times $\nu_q$ and $\nu_g$ are not equal, i.e. $\nu_q \neq \nu_g$. In
fact, for $\bar{\gamma} \gg 1$ the contributions of the terms
$\nu_q/\bar{\gamma}$ and $\nu_g/\bar{\gamma}$ are much less than unity
and can be neglected as well as $\nu/\bar{\gamma}$. Thus, the
ultra--relativistic and spherical symmetric expanding QGP behaves like
a collisionless thermalized plasma.

\section{Conclusion}
\setcounter{equation}{0}

\hspace{0.2in} Within Quark--gluon transport theory outlined in
Refs.[2,3] we have investigated the contributions of a chemical
potential of light quarks and antiquarks $\mu(T)$ and a non--zero
value of a mass $m_s$ of strange quarks and antiquarks to the
chromoelectric permeability of the QGP at rest and the
ultra--relativistic and spherical symmetric expanding.

For the QGP at rest we have shown that the contribution of a chemical
potential of light (massless) $u$ and $d$ quarks and antiquarks does
not affect the formation of the chromoelectric permeability of the QGP
and can be neglected. In turn, the account for a non--zero value of a
strange quark mass $m_s$ has led to the chromoelectric permeability
which might be produced in the QGP constituted with light $u$ and $d$
quarks, $\bar{u}$ and $\bar{d}$ antiquarks and gluons only. The
numerical estimate, carried out for $m_s = 135\,{\rm MeV}$ and $T =
175\,{\rm MeV}$, a typical freeze--out temperature of the QGP produced
in ultra--relativistic heavy--ion collisions, has shown almost
complete screening of massive strange quarks and antiquarks for the
formation of the chromoelectric permeability of the QGP.  Such a {\it
decoupling} of massive strange quarks and antiquarks from the
quark--gluon system can be expressed in terms of a plasma frequency
$\omega^2_0$ decreased from the value $\omega^2_0 = g^2_sT^2(2N_C +
N_F)/3$ calculated in Refs. [2,3] for massless quarks and antiquarks
at $N_F = 3$, the number of quark flavours, to the value $\omega^2_0 =
g^2_sT^2(2N_C + N_F - 1)/3$ obtained in this paper. This means that
massive strange quarks and antiquarks are not material for
oscillations and instabilities which can be induced in the QGP [2,3]
at the freeze--out temperature $T = 175\,{\rm MeV}$.

The extension of the results obtained for the QGP at rest to the case
of the ultra--relativistic and spherical symmetric expanding QGP,
which can be produced in ultra--relativistic heavy--ion collisions in
the center of mass frame of colliding ions, has shown that the
ultra--relativistic and spherical symmetric expanding QGP behaves like
a collisionless thermalized plasma. We have found that in the
ultra--relativistic and spherical symmetric expanding QGP the plasma
frequency $\omega_0$ becomes enhanced by a factor $\bar{\gamma}\simeq
2\,\gamma$, i.e. $\bar{\omega}_0 \simeq 2 \gamma \omega_0$, where
$\gamma =1/\sqrt{1 - v^2}$ is a Lorentz factor of the
ultra--relativistically moving QGP with a hydrodynamical 3--velocity
$v \sim 1$, relative to the plasma frequency of the QGP at rest. Since
for the ultra--relativistic QGP $\gamma \gg 1$, the plasma frequency of
the ultra--relativistic and spherical symmetric expanding QGP is much
greater than the plasma frequency of the QGP at rest $\bar{\omega}_0
\gg \omega_0$.

We would like to accentuate that the expressions for the
chromoelectric permeability of the ultra--relativistic and spherical
symmetric expanding QGP as well as for the QGP at rest, calculated in
our paper, agree with the chromoelectric permeability tensor
calculated in Refs.[2,3] for the QGP constituted with massless $u,d,s$
quarks, antiquarks and gluons. The results are in analytical agreement
up to the replacement of the plasma frequency. This is $\omega^2_0 =
g^2_sT^2(2N_C + N_F)/3 \to \omega^2_0 = g^2_sT^2(2N_C + N_F - 1)/3$
for the plasma at rest and $\omega^2_0 = g^2_sT^2(2N_C + N_F)/3 \to
\bar{\omega}^2_0 = 4\gamma^2 g^2_sT^2(2N_C + N_F - 1)/3$ for the
ultra--relativistic plasma. This means that all dispersion relations
derived in Ref.[2,3] are valid in our case. However, due to the
enhancement of the plasma frequency $\bar{\omega}_0 \gg \omega_0$,
caused by a relativistic motion of the QGP, all singularities of the
QGP at rest obtained in Ref.[2,3], should be shifted to the region of
very high frequencies in the case of the ultra--relativistic and
spherical symmetric expanding QGP.

\end{document}